# Leveraging Wireless Sensor Networks for Real-Time Monitoring and Control of Industrial Environments

Muhammad Junaid Asif [1,3*], Abdul Rehman[2#], Asim Mehmood[3], Muhammad Hamza[4], Rana Fayyaz Ahmad [1], Shazia Saqib[5]

[1] Artificial Intelligence Technology Centre (AITeC), National Centre for Physics (NCP) Islamabad 44000, Pakistan
[2] Department of Electrical Engineering, University of Engineering and Technology (UET) Lahore 54890, Pakistan
[3] Faculty of IT and Computer Sciences (FoIT&CS), University of Central Punjab (UCP) Lahore 53700, Pakistan
[4] School of Engineering and Energy, Murdoch University, Perth 6150, Australia
[5] School of Informatics and Robotics, Institute of Arts and Culture (IAC), Lahore Pakistan

*junaid.asif@ncp.edu.pk; #2019ee78@student.uet.edu.pk

***Abstract*— This research proposes an extensive technique for monitoring and controlling the industrial parameters using Internet of Things (IoT) technology based on wireless communication. We propose a system based on NRF transceivers to establish a strong Wireless Sensor Network (WSN), enabling transfer of real-time data from multiple sensors to a central setup that is driven by Arduino microcontrollers. Different key parameters, crucial for industrial setup, such as temperature, humidity, soil moisture and fire detection, are monitored and displayed on an LCD screen, enabling factory administration to remotely oversee the industrial operations remotely over the internet. Our proposed system eliminates the need for physical presence during monitoring by addressing the shortcomings of conventional wired communication systems. In addition to monitoring, the system includes an additional feature to remotely control these parameters by controlling the speed of DC motors through online commands. Given the rising incidence of industrial fires worldwide between 2020 and 2024 due to an array of hazards, this system with dual functionality boosts the overall operational efficiency and safety. This overall integration of IoT and Wireless Sensor Network (WSN) reduces the potential risks linked with physical monitoring, providing rapid responses in emergency scenarios, including the activation of firefighting equipment. The results show that innovations in wireless communication play an integral part in industrial process of automation and safety, paving the way to more intelligent and responsive operating environments. Overall, this study highlights the potential for change of IoT-enabled systems to revolutionize monitoring and control in a variety of industrial applications, resulting in increased productivity and safety.**

***Keywords*— Internet of Things (IoT), Industrial parameters monitoring, Wireless Sensor Network (WSN), Soil moisture measurement, Temperature measurement, Fire detection.**

## I. INTRODUCTION

The field of Artificial Intelligence (AI) has seen remarkable progress, with the emergence of the Internet of Things (IoT) standing out as a major technological advancement in the era of wireless communication and mobility. This innovation has permeated various sectors such as manufacturing, supply chain management, industrial operations, warehousing, and healthcare by integrating sensors and actuators to enhance overall efficiency. Despite advancements in traditional wired communication systems, many industrial operations still rely on physical presence for monitoring and control, leading to inefficiency, safety concerns, and limited operational flexibility. IoT offers a compelling solution by enabling remote monitoring and control, eliminating the need for continuous physical supervision. This shift can lead to increased productivity, enhanced safety protocols, and a more agile approach to industrial processes. In an industrial environment with multiple machines, constant human presence is necessary to monitor and control various parameters like temperature, speed, and torque. Manual supervision of these processes is time-intensive, costly and prone to data loss and manipulation.

A significant rise in fire and explosion incidents has been observed in Pakistan from 2020 to 2022, likely attributable to various technical factors such as electrical short circuits, smoking, and the presence of flammable liquids and gases. These incidents are not confined to industrial settings but are occurring across commercial, residential, and agricultural sectors. Traditional fire alarm systems may not be sufficiently advanced to address these challenges. Implementing IoT-based intelligent fire detection systems can enhance the ability to swiftly detect fires and trigger essential actions such as cutting off the electricity supply and activating water sprinklers, all without the need for immediate physical intervention.

Currently, Internet connectivity is widespread and readily available, enabling rapid and dependable monitoring and information sharing [1]. The Internet of Things (IoT) encompasses a network of physical devices equipped with electronics, software, sensors, and connectivity to gather and exchange data efficiently [2], [3]. IoT is proving to be crucial across various sectors such as automotive, healthcare, transportation, agriculture, automation, and smart cities [4].

Wireless Sensor Networks (WSN) consist of individual sensor-equipped nodes strategically positioned to monitor environmental and physical variables collaboratively. These devices work together to observe parameters such as temperature, motion, and location [5]. WSN facilitate multi-hop communication and enables diverse network topologies. Recent progress in wireless communication technologies and integrated circuits has made it possible to develop low-power, cost-effective, and efficient devices suitable for WSN applications [6].

This paper introduces an innovative approach that integrates IoT and WSN to oversee and regulate various industrial parameters such as temperature, humidity, soil moisture, and flame detection. By utilizing these parameters, users can swiftly take the required actions to remotely operate relevant machinery through actuators via the Internet, eliminating the need for physical presence. Initially, various sensors will be strategically placed at specific locations to capture relevant data, which will then be transmitted to the microcontroller through a wireless sensor network. The collected data will be stored and sent to a monitoring and control unit via the Internet. This setup allows individuals to remotely monitor and manage different parameters without the need for any physical intervention. The main contributions of this research are summarized as follows:

- Integration of IoT and WSN technologies for real-time and remote monitoring and control of industrial parameters via the Internet using the low-cost ARDUINO Uno/Nano nodes and nRF24L01 transceivers, enabling rapid and accurate responses without physical intervention.
- A bi-directional IoT gateway is designed to compile multisensory data and give instructions to actuators via a secure internet interface thereby enabling closed-loop monitoring and control.
- Functional performance was evaluated using various evaluation metrics such as latency, packet-delivery rate (PDR), and reliability.

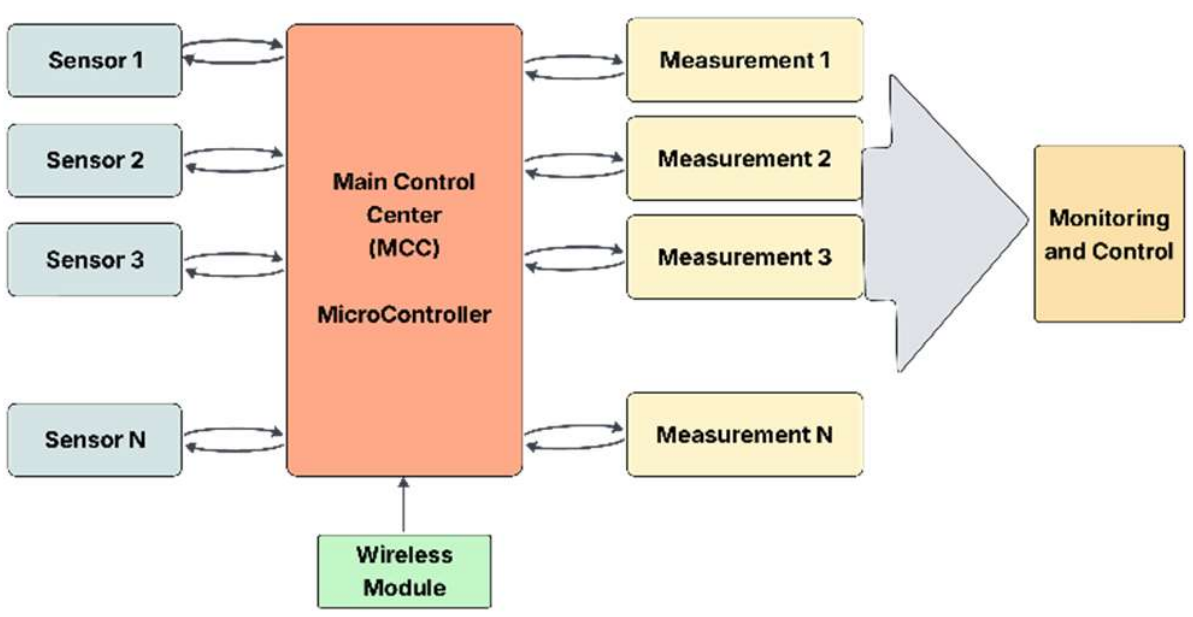

**Fig. 1.** Overview of the proposed IoT and WSN-based system

The paper is structured as follows: Section II presents a comprehensive literature review, Section III outlines the detailed methodology, Section IV describes the system implementation, Section V presents the results, and Section VI provides concluding remarks.

## II. LITERATURE REVIEW

The field of the Internet of Things (IoT) and industrial automation is witnessing significant research and development in Pakistan. Various research teams from different universities are partnering with industrial partners to introduce automation in industrial processes. This trend is swiftly gaining momentum across multiple sectors such as agriculture, power, healthcare, home automation, and renewable energy.

An IoT-and Raspberry Pi-based system was developed to monitor various industrial processes. Diverse sensors were employed to track parameters such as temperature, light intensity, water level, current, and voltage across the industrial workflow. Real-time data was monitored, recorded, and stored in the cloud and made accessible via smartphones and computers. The key feature of proposed system was the ability to monitor multiple factors remotely through the internet [7].

An IoT-based condition and availability monitoring system was proposed for real-time monitoring of CNC machines. The system aimed to perform proactive maintenance of the CNC machines daily by observing the temperature, vibration, current, and lubricant level of the machine. All these parameters were monitored, recorded, and displayed on a cloud-based platform, facilitating rapid fault diagnosis and minimizing machine downtime regardless of the cause [8].

Numerous researchers in agricultural engineering are actively engaged in creating IoT-based systems to automate agricultural practices. The agricultural sector holds significant economic importance for any nation as it directly impacts production capacity and the country's trade activities. A solution based on IoT approach was proposed to manage and control insecticides, pesticides, water levels, and crop management in agricultural setting [9], [10].

Another study focused on the development of AGRIBOTS using IoT technology to oversee and regulate various water irrigation parameters. Key parameters under surveillance and control encompass drip irrigation, temperature, water levels, and crop growth [11].

A framework was introduced that utilizes a variety of sensors to detect fire, smoke, moisture, temperature, current, and voltage. This framework is designed to gather, analyze, and interpret data from machines, engines, and devices employed in industrial environments [12], [13].

An intelligent system for monitoring and controlling agricultural greenhouses was developed by integrating IoT and WSAN technologies. The system was created using Python programming language, a Human-Machine Interface (HMI), and the Node-RED Server by IBM, offering a comprehensive solution for efficient greenhouse management [14].

A system for automated monitoring of air quality (AQ) and LPG leaks IoT-based technology was introduced. A Support Vector Machine-driven system was recommended for monitoring and alerting indoor air quality levels, aiming to mitigate health hazards [15].

## III. METHODOLOGY

The research incorporates Arduino Uno R3 [16], [17] and the Arduino Nano as central modules for managing connected components and enabling data collection and internet communication. Wireless connectivity is achieved through the nRF24L01 module, facilitating communication between the Arduino Nano and Arduino Uno microcontrollers, as depicted in Figure 2.

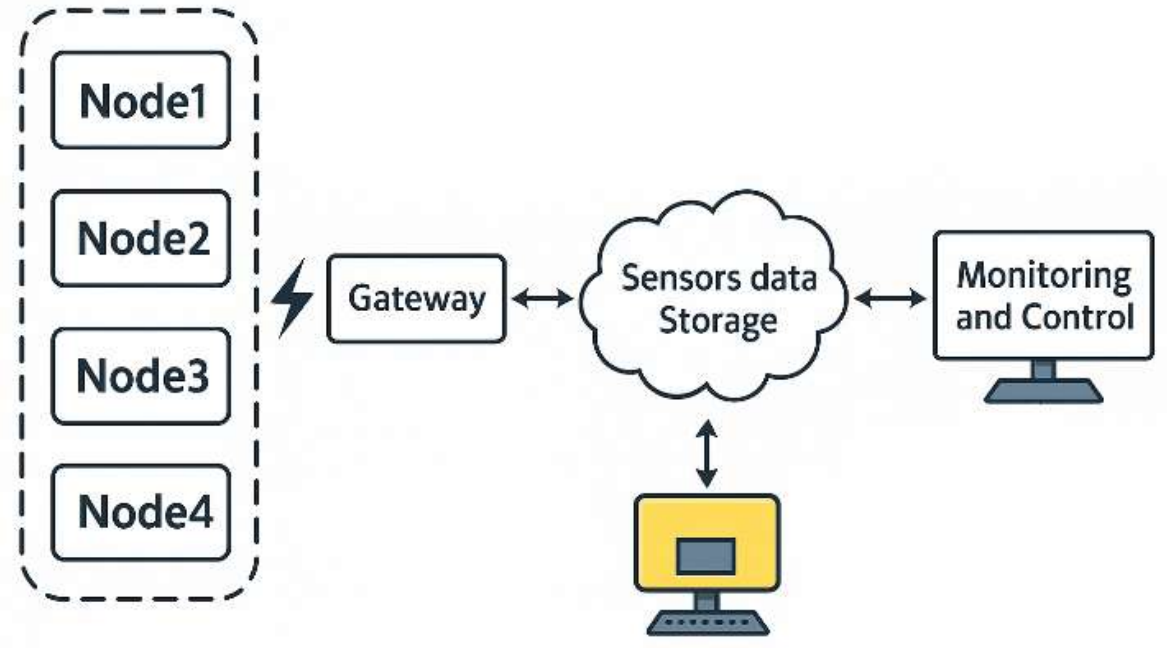

**Fig. 2.** Sensor network architecture showing data flow from distributed nodes to centralized storage via a gateway.

The system integrates IR flame, soil moisture, temperature, and humidity sensors, alongside remote monitoring and control of the DC motor's speed via the internet. The system's operational flow is visually represented in ***Fig. 3***. Powering the system uses a USB hub connected to the main adapter. Each node receives power individually through USB cables, thereby initializing power system. Nodes then initiate data collection from sensors for transmission to the gateway using nRF24L01 wireless communication modules.

Data received at the gateway is visualized and transmitted to the internet through an Ethernet shield. The bidirectional arrows in the flowchart symbolize the data exchange, illustrating data transmission to the internet and receipt of commands for regulating the DC motor.

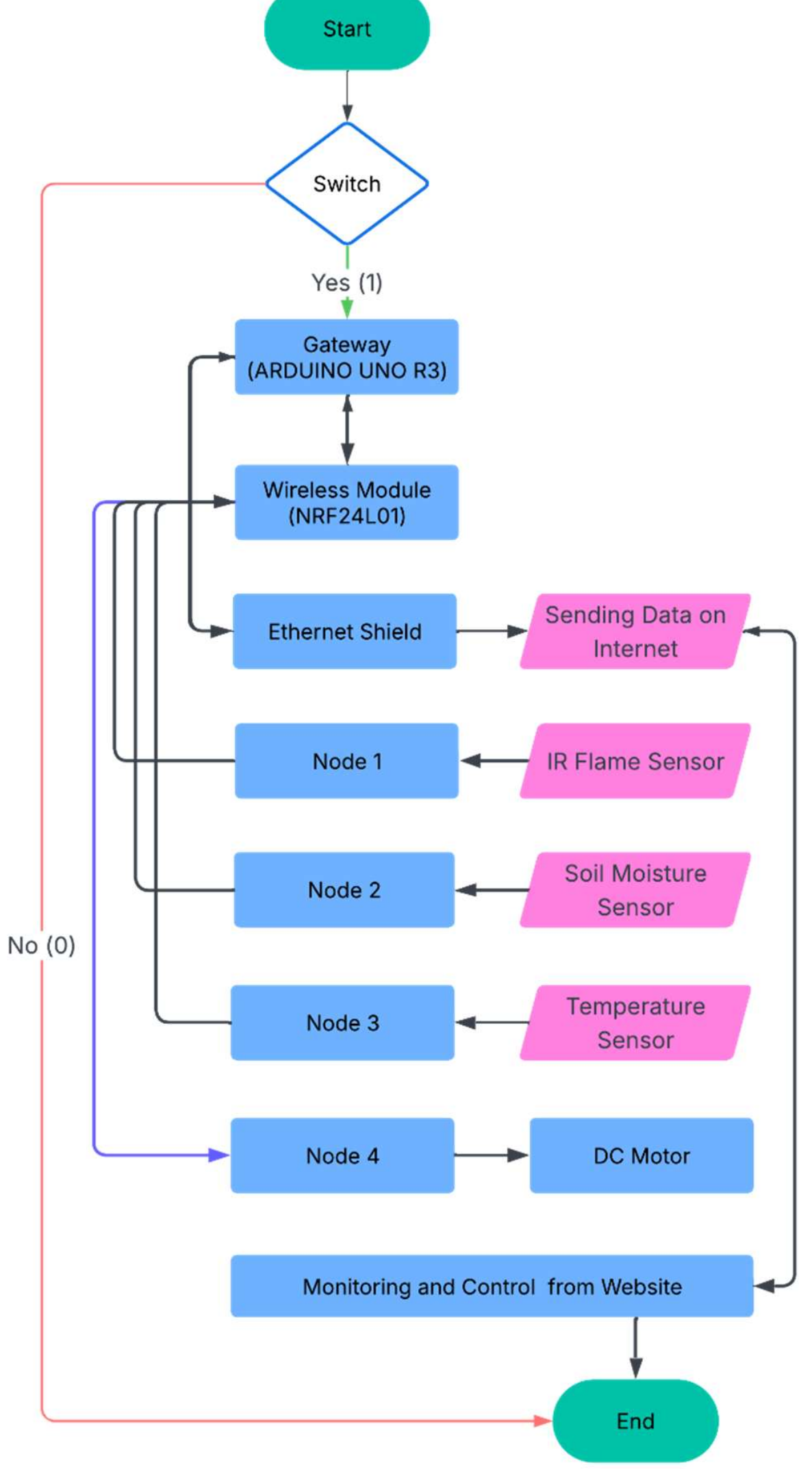

**Fig. 3.** Flowchart illustrating the operational workflow of the proposed system.

## IV. IMPLEMENTATION

### *A. System Working Methodology:*

In addition to the sensor components, the nRF24L01 module is a critical element of the sensor node. This module comes equipped with a built-in PCB antenna and functions as an ultra-low power transceiver, consuming 12mA during transmission and operating at a 2.4 GHz carrier frequency. It supports baud rates ranging from 250kbps up to 2Mbps and can accommodate up to 125 independent modems using 125 distinct channels [18], [19].

Every sensor element is directly connected to the Arduino Nano, which is powered by the ATmega328P microcontroller. With an operating voltage of 5V, the Arduino Nano features 14 digital I/O pins, including 6 that support PWM output. It is equipped with 8 analog pins, and each I/O pin can handle a current of up to 40 mA. The Arduino Nano has a flash memory limit of 16KB and an SRAM capacity of 1KB as specified in [20], [21].

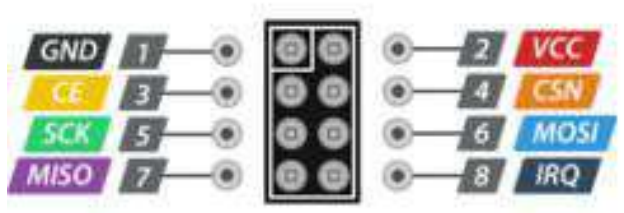

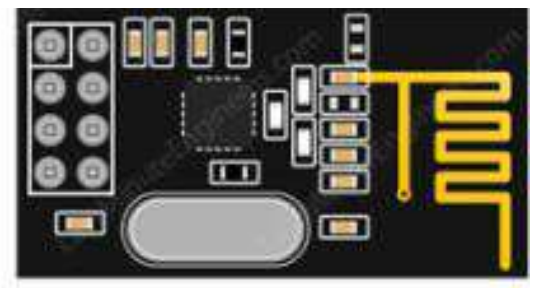

**Fig. 4.** Detailed Pin Configuration of the NRF24L01 Wireless Transceiver Module

### *B. Sensor Nodes:*

Each sensor is individually connected to an Arduino Nano, creating distinct nodes such as ***Node 1, Node 2, Node 3, and Node 4***, each responsible for different sensor functionalities. The sensor data from each node is transmitted to the Arduino Uno via nRF24L01 modules and subsequently forwarded to the internet through an Arduino Ethernet Shield.

- **Node 1** consists of an IR flame sensor, nRF24L01, and ARDUINO Nano. The IR flame sensor features various flame detection methods such as UV and IR detectors. It is powered by connecting its Vcc (pin 3) to the Arduino 5V pin and analog pin A0 to the sensor's pin 1.
- **Node 2** is centered around a soil moisture sensor, nRF24L01, and Arduino Nano. The sensor triggers a digital output when the measured electrical resistance between its contacts surpasses a set threshold. The Vcc pin of the sensor is connected to the Arduino's 5V pin, and Arduino's analog pin A0.
- **Node 3** includes a DHT11 temperature and humidity sensor, nRF24L01, and Arduino Nano. The sensor's data pin 1 is connected to Analog pin A0 of Arduino Nano, while the 5V pin is connected to the sensor's Vcc. The DHT11 provides temperature readings with an accuracy of $\pm 1$°C and humidity readings with an accuracy of $\pm 1\%$.
- **Node 4** comprises a DC motor, L298N motor driver, Arduino Nano, and nRF24L01 module. The DC motor, operating at 5V, is connected to the Nano via the L298N motor driver. The L298N driver allows control over two DC motors simultaneously in terms of speed and direction, with a maximum current capacity of up to 2A and compatibility with DC motors ranging from 5V to 35V.

The Nano connections are established such that pin 8 and pin 9 of the Nano are connected to the IN3 and IN4 pins of the L298N. The output pins 3 and 4 of the L298N are connected to the DC motor for operational control.

### *C. Gateway:*

The gateway system comprises an Arduino Uno, an Ethernet shield, an I2C module, nRF24L01, and 16 x 4 LCD modules. The Arduino Uno is a development board featuring the ATmega328P microcontroller, with 14 digital I/O pins, a 16 MHz crystal oscillator, 6 analog pins, a USB port, an ICSP header, and a reset button. It comes with a pre-loaded bootloader, eliminating the need for additional hardware during programming. Among the 14 digital pins, 6 can function as PWM outputs.

The primary function of the gateway is to receive sensor data, display it on the LCD modules, and transmit the data to the internet. The gateway operates in a bidirectional communication mode, receiving data from the nodes and sending command signals to control the DC motor. The nRF24L01 module utilizes the SPI protocol, MISO and MOSI pins. For the nRF24L01 node to function as a transmitter, MOSI is set to 1 and MISO is set to 0, and vice versa for receiver mode. All sensor nodes transmit data following a cluster tree topology. The router connects to the gateway through the Ethernet shield, facilitating data exchange, as illustrated in ***Fig. 05.***

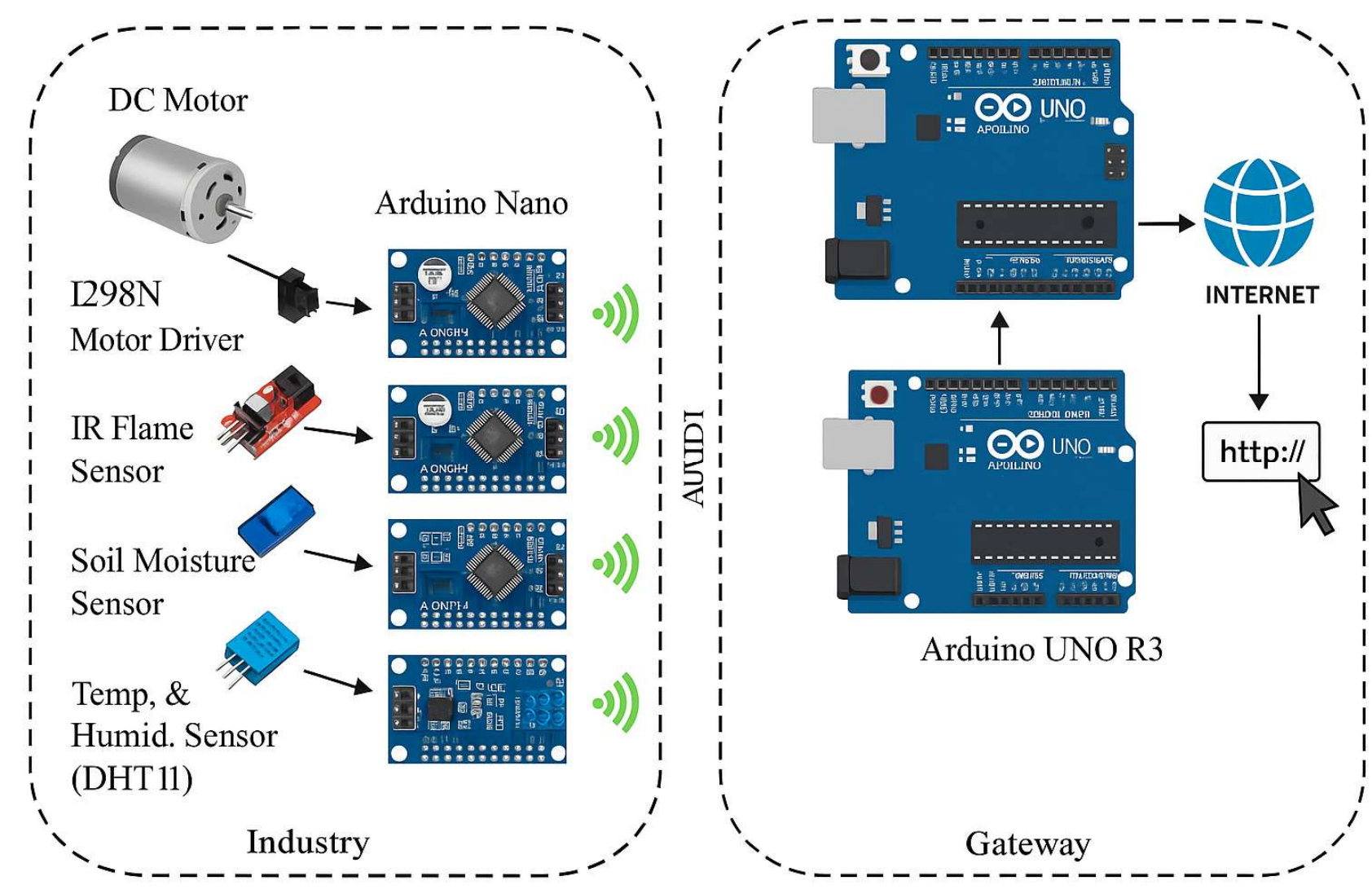

**Fig. 5.** System architecture showing key components and data flow between sensors, processing units, and the gateway.

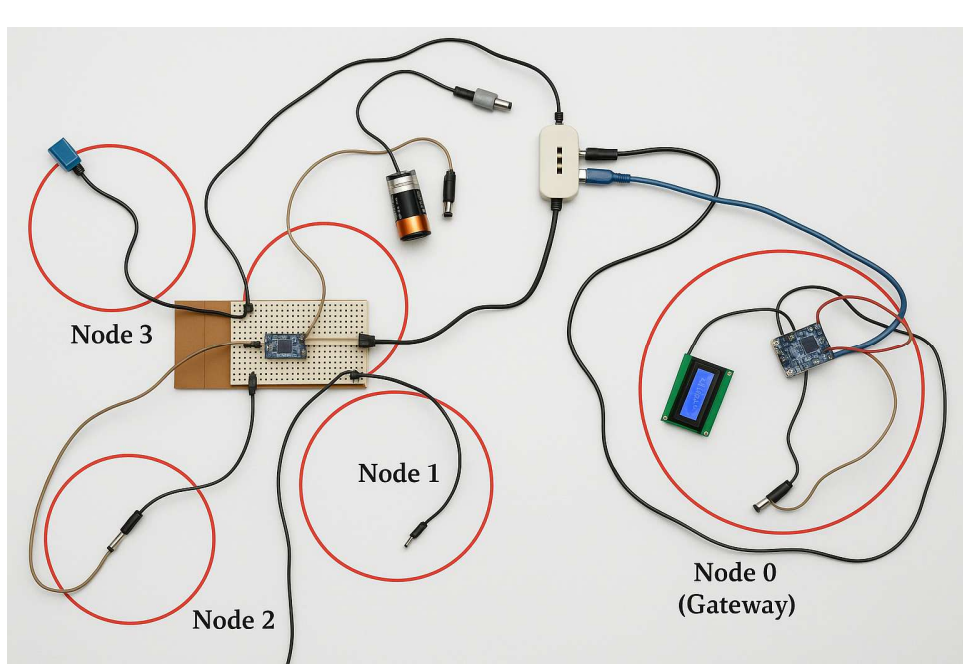

**Fig. 6.** Hardware deployment showing interfacing of sensors with the IoT gateway

The I2C module is used in this research project to reduce the number of LCD pins from 16 to 3. These three pins are SDA, SCL and GND and they enable synchronous data transfer from the gateway to LCD. Due to the limited number of pins on Arduino, the I2C module is integrated with the LCD to support its functionality, as it requires multiple pins. By connecting the LCD with I2C, the data can be displayed with the help of three pins.

To change the motor direction, the user can simply change the rotation from the internet through a website, the data from the website will be sent to the gateway Arduino Uno R3, from there it will be transmitted back through nRF24L01 module to the Arduino Nano (Node4) on which the motor is connected, from there Nano will change the rotation of the DC motor. As node4 contains, L298N and nRF24L01 connected to the Arduino Nano. The functionality of this node is to receive a value from the gateway Arduino and then set the speed of motor which is connected to the L298N accordingly.

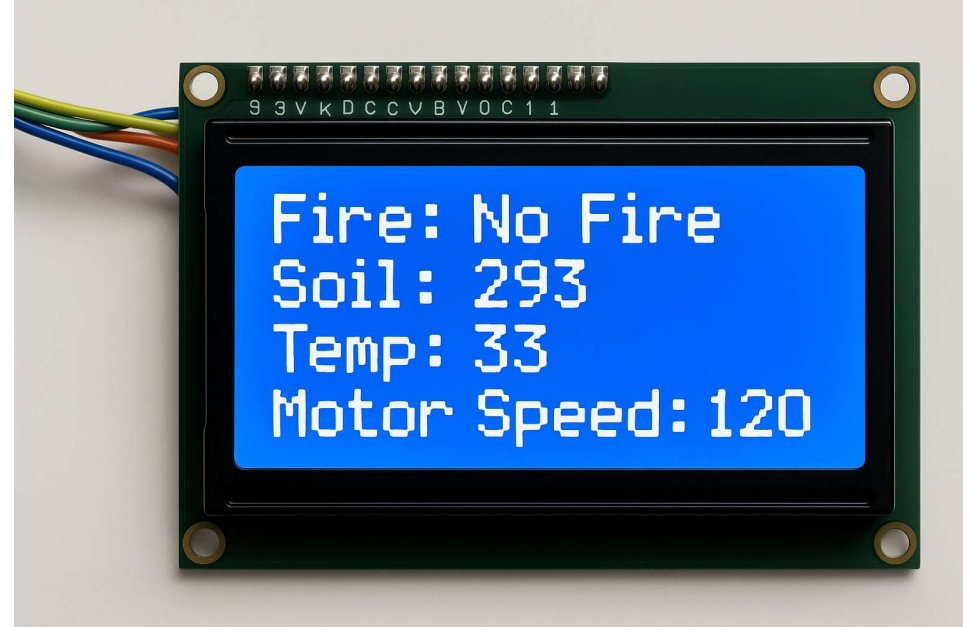

**Fig. 7**. Real-Time Sensor Data Visualization Interface on IoT Gateway Device

### *D. Error Analysis and Fault Tolerance*

To achieve the desired, industrial-level reliability, the proposed system includes multiple tiers of fault detection, fault analysis, and recovery.

- **Sensor noise and drift:** Signals from time-series anomaly detection by rolling mean and standard deviation assessment. When the values are more than three standard deviations, it raises outliers.
- **Communication Errors:** This is detected by using the packet sequence numbers, cyclic redundancy check (CRC) responses, as well as the acknowledgment (ACK) responses between the nodes and the gateway.

- **Node or gateway Failure:** It is detected by non-reception of signal messages or excessive time inactivity.
- **Data Inconsistency:** It was identified through cross-checking sensor data patterns and correlation amongst multiple nodes.

### *E. Performance Metrics:*

To evaluate the reliability and operational stability of the proposed IoT -WSN framework completely, four main performance indicators have been evaluated:

- Packet Delivery Ratio (PDR),
- Mean Time Between Failures (MTBF),
- Latency [22], and
- False Alarm Rate

The PDR [23] is the percentage of data packets received by the receiver of the channel compared to the total packet sent by the transmitter indicating the efficiency of the wireless communication link and reliability of the network. Over the seven days of evaluation period, the system had an average PDR of 96.4 meaning that there was a slight loss of packets even when the environment was interfering.

MTBF [24] is the average of the time between failures of the system or node and is used to measure the strength of the hardware and the communication of a system or a node; the mean time was found to be around 120 hours.

Taken together, all these metrics show that the system has high reliability, fast fault recovery, and high sensing accuracy, which confirms its usability in the context of continuous industrial monitoring applications.

### *F. Security Considerations*

Since the system was interconnected with the wireless nodes and the internet, it was designed with several layers of security to ensure data integrity and confidentiality.

**Wireless Link Security:**

- The firmware has AES-128 encryption that is employed to send messages over nRF24L01.
- A 16-byte authentication tag (HMAC-SHA256) is added to every message to prevent replay or tampering attacks.
- It eliminates duplicate packets and out of order packets by using sequence numbers and timestamps.
- Each node is pre-configured with unique encryption keys; it is possible to add key rotation in the future.

**Gateway to Cloud Security:**

- The gateway sends the information to the cloud server via HTTPS (TLS 1.2 or more) or MQTT over TLS to transfer the encrypted information.
- Ratio authentication of certificates is done to server and API keys with limited permissions are applied.

**Physical and Operational Security.**

- Access to nodes is limited physically.
- Post programming Disables debugging ports.

## V. Results

In this section, the results pertaining to temperature, soil moisture, and flame sensors are presented. Flame sensor measurements are depicted in Fig. 8(a). The plot shows the flame detected by the sensor from a distance. Conversely, in the absence of a flame, the sensor registers a value of 0. Fig. 8(b) illustrates the progression of temperature and humidity values recorded until 2:30PM. Furthermore, Fig. 8(c) illustrates the soil moisture data.

The data transmitted by the sensor nodes and received on the web is imported to create datasets. The system's operational and analysis timeframe is set from 10:30 AM to 2:30 PM, with parameter values recorded at 30-minute intervals.

Table 2 presents a sample of the data received at the gateway, which is displayed on the Arduino UNO console. Soil moisture values are gathered by placing the sensor in soil over the specified time frame. Higher values indicate increased moisture content, such as water in the soil, while lower values signify drier conditions as the sensor is transferred to a dry medium. For the flame sensor, values are acquired by introducing a flame close to the sensor.

**Table I: Evaluation of proposed model using performance metrics**

| Metric | Symbol | Measured Value (for 4 hours Testing) | Measured Value (for 7 days Testing) | Acceptance Criterion |
|---|---|---|---|---|
| **Packet Delivery Ratio** | **PDR** | **96.4 %** | **95.7%** | **≥ 95 %** |
| **Median Latency** | $L_{50}$ | **180 ms** | **220 ms** | **< 200 ms** |
| **95th Percentile Latency** | $L_{95}$ | **900 ms** | **950ms** | **< 1 s** |
| **False Alarm Rate (fire)** | **FAR** | **0 %** | **0.2%** | **≤ 1 %** |
| **MTBF** | **–** | **120 h** | **108h** | **≥ 100 h** |
| **Power per Node** | **P** | **42 mA avg** | **44mA avg** | **≤ 50 mA** |

**Table II: Environmental parameter readings from wireless sensor node**

| Parameter | Value |
|---|---|
| **Date** | 09-01-2025 |
| **Time** | 10:28 |
| **Temperature (°C)** | 33.00 |
| **Soil Moisture (ADC Value)** | 293 |
| **Humidity (%)** | 70.00 |

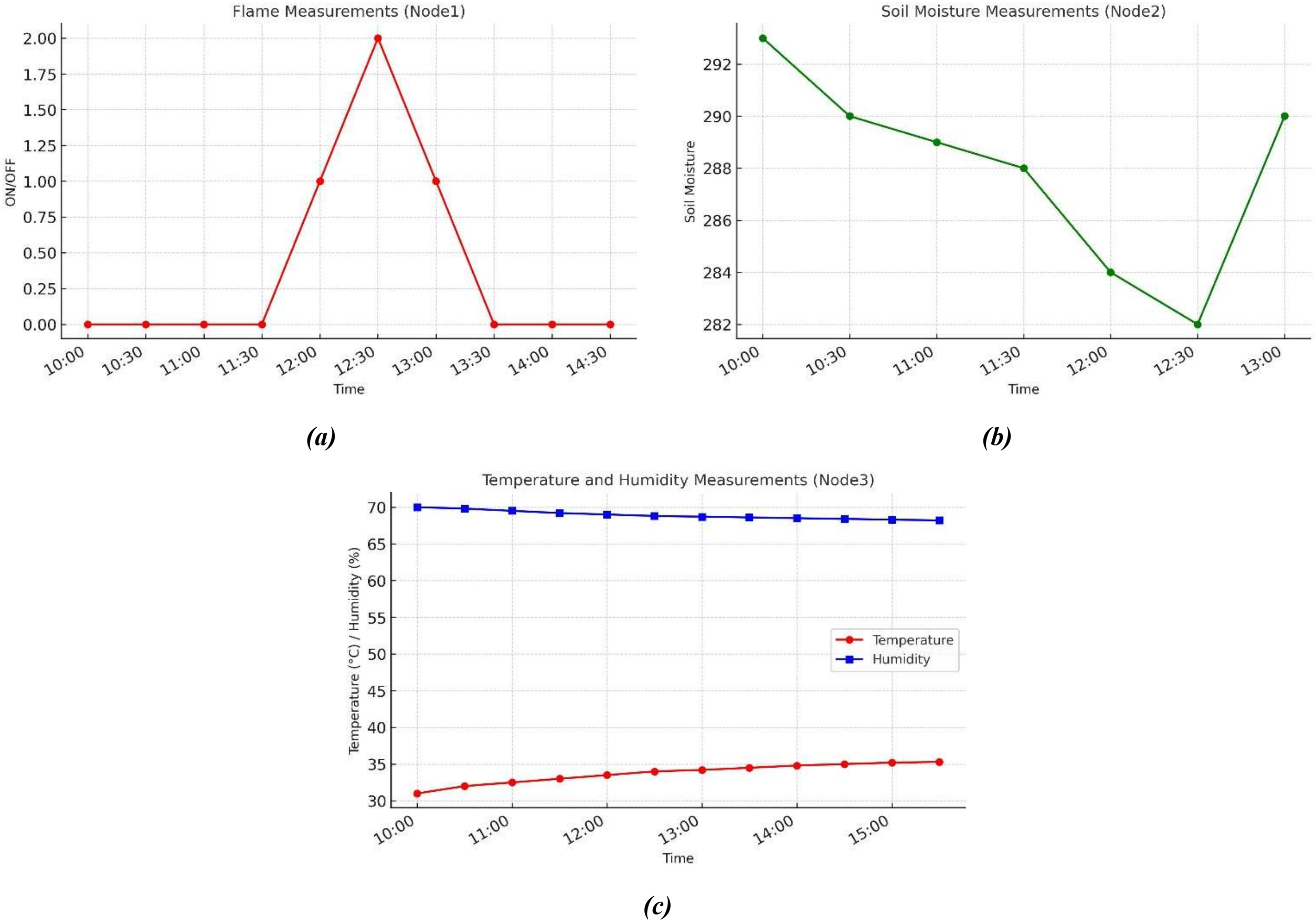

**Fig. 08**. ***(a).*** Flame detection measurements from Node 1 between 10:00 and 14:30. The sensor recorded no flame activity for most of the period, except between 12:00 and 12:30, where a peak value indicates flame presence, ***(b).*** Soil moisture measurements recorded by Node 2 over the observation period from 10:00 to 14:30. The graph shows a gradual decline in soil moisture until 12:30, reaching a minimum around 13:00, followed by a recovery towards 14:30, ***(c).*** Temperature and Humidity measurements recorded by Node 3 over the observation period from 10:00 to 14:30.

### *A. Extended Reliability and Performance Testing*

The prototype was tested to ensure long-term stability and industrial preparedness, which is why it was tested on long-term operation after the initial testing of 4 hours. Overall results are as shown in Table 1.

**Test Configurations:**

- **Time span:** 7 days (168 hours) on the air.
- **Configuration:** 4 sensor nodes (temperature, soil moisture, flame, and one actuator node) were linked to one gateway.
- **Time rate of data transmission:** 30 seconds per node.

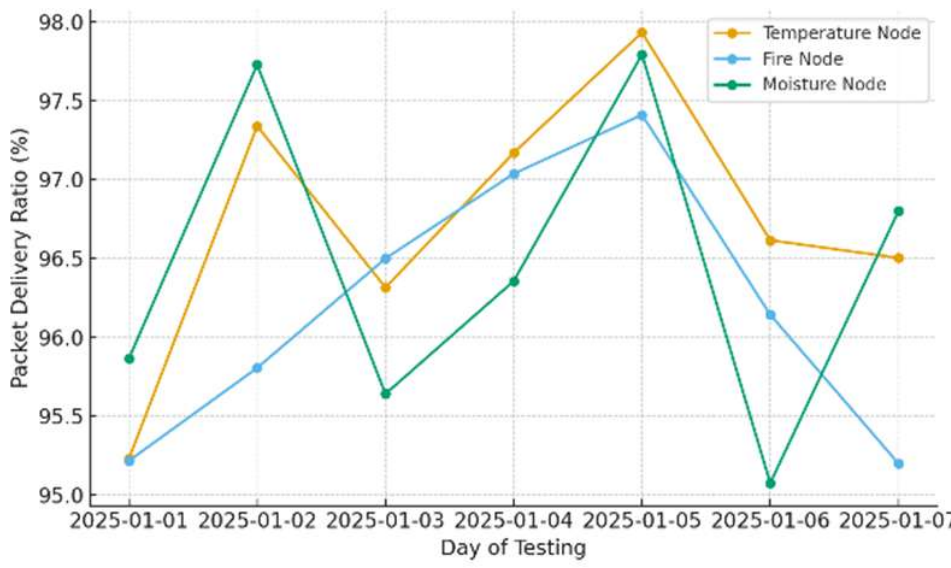

**Fig. 09**. Extended seven-day reliability test showing PDR for three sensor nodes.

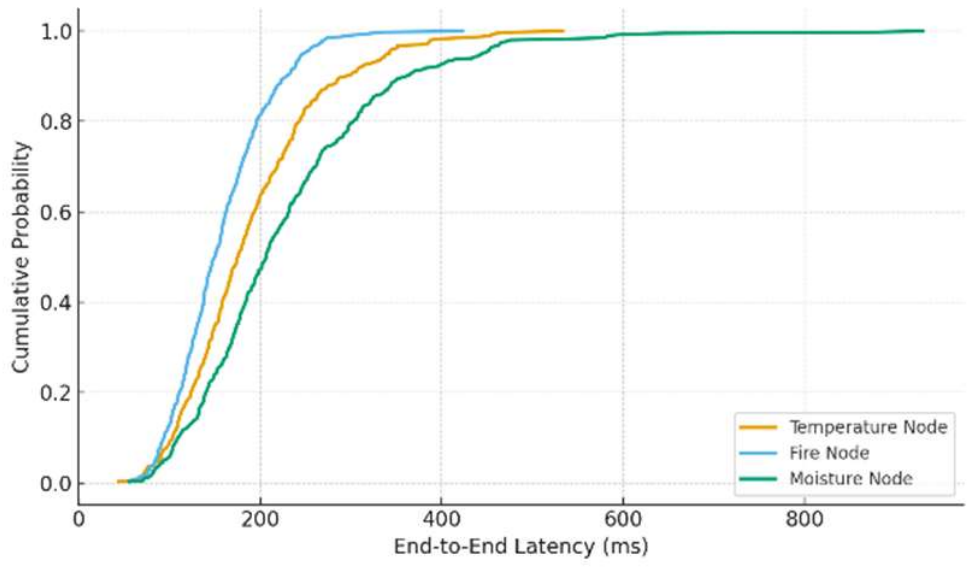

**Fig. 10**. CDF for end-to-end latency of three sensor nodes

## VI. CONCLUSIONS

In conclusion, the implementation of the Wireless Sensor Network (WSN) was deemed successful, with data transmission to the gateway achieved through a wireless medium. NRF24L01 served as the wireless transceiver for sending and receiving data in this study.

Upon integrating all components and varying parameters across each node, the accuracy of values displayed on the gateway node indicated proper functionality of the setup. Subsequently, the values from each node were forwarded to the THINGSPEAK website. Sensor values were consistently and accurately displayed at defined intervals, and the control of the motor speed was effectively managed. This underscores the potential utility of WSN and IoT technologies for comprehensive wireless monitoring and device control in an increasingly evolving world.

### *A. Future Work:*

This framework is specifically designed to solve the issue of thermal safety and motor-speed control in small industrial settings e.g. furnace zones and paint booths. The system can identify any temperature changes or fire incidences and automatically cut or switch motors remotely.

The future improvements involve:

- **Predictive Maintenance:** On-board machine learning Predictive maintenance involves the installation of vibration sensors to alert mechanical defects before the machine breaks.
- **Adaptive Sensing:** Starts sampling at varying rates depending on the anomaly noticed to optimize power and performance.
- **Edge Intelligence:** Incorporating simple anomaly detection on the gateway to decrease reliance on the cloud and decrease latency.
- **Field Deployment**: Large scale tests are also to be conducted in actual industrial equipment, to determine integrity of EMI, and compatibility with PLC/SCADA.